\documentclass[prl,preprint,groupedaddress,showpacs]{revtex4-1}

\usepackage{amsmath}
\usepackage{amssymb}
\usepackage{amsfonts}
\usepackage{mathrsfs}
\usepackage{epsfig}
\usepackage{bm}
\usepackage{multirow}
\usepackage{color}
\usepackage{lineno}

\DeclareMathOperator{\sign}{sign}

\begin{document}

\title{Optical control of the topology of laser-plasma accelerators}
%\title{Twisted plasma acceleration}
\author{J. Vieira$^{1}$}
\author{J.T. Mendon\c ca$^1$}
\author{F. Qu\'er\'e$^2$}

\affiliation{$^1$GoLP/Instituto de Plasmas e Fus\~{a}o Nuclear, Instituto Superior T\'{e}cnico, Universidade de Lisboa, Lisbon, Portugal}
\affiliation{$^2$LIDYL, CEA, CNRS, Universit\'e Paris-Saclay, CEA Saclay, 91 191 Gif-sur-Yvette, France}
\today

\begin{abstract}

We %demonstrate that %laser-driven 
%plasma based accelerators can provide unprecedented flexibility on the topology of the accelerating structure, 
propose a twisted plasma accelerator capable of generating relativistic electron vortex beams with helical current profiles. The angular momentum of these vortex bunches is quantized, dominates their transverse motion, and results in spiraling particle trajectories around the twisted wakefield. We focus on a laser wakefield acceleration scenario, driven by a laser beam with a helical spatiotemporal intensity profile, also known as a light spring. We find that these light springs can rotate as they excite the twisted plasma wakefield, providing a new mechanism to control the twisted wakefield phase velocity and enhance energy gain and trapping efficiency beyond planar wakefields.
\end{abstract}

\maketitle

Using %lasers 
plasmas
to accelerate particles to high energies has long been identified as a promising path to obtain compact accelerators~\cite{bib:tajima_prl_1979,bib:modena_nature_1995,bib:malka_science_2002}. In terms of particle energy, the most advanced scheme to date for electrons consists in using ultraintense femtosecond lasers %
or electron beam drivers %
to excite high-amplitude ultra-relativistic waves in low-density plasmas~\cite{bib:pukhov_apl_2002,bib:lu_prl_2007, bib:lu_prstab_2007,bib:esarey_rmp_2009}. Electrons trapped into these waves can %then
gain energy and be accelerated to relativistic velocities~\cite{bib:mangles_nat_2004,bib:geddes_nat_2004,bib:faure_nat_2004,bib:kim,bib:leemans_prl_2014}. 

%Current approaches successfully control the longitudinal phase-space of the accelerated bunch (e.g. longitudinal momentum). Manipulating the transverse beam structure is much more challenging. This limitation is intrinsically related to the plasma wave structure. 

%Plasma waves are central in plasma acceleration, and any fundamental change to the plasma wave structure will then have deep ramifications into the physics and potential applications of plasma accelerators.

Currently, longitudinal phase-space properties of the accelerated bunches, such as their longitudinal momentum, can be effectively controlled. Despite recent progresses to control the radial dynamics of accelerated beams, using sophisticated temporal wakefield modulations%(e.g. oscillating plasma waves leading to undulator radiation
~\cite{bib:rykovanov_prl_2015} 
and spatial shapes %(e.g. doughnut plasma waves for positron acceleration
~\cite{bib:vieira_prl_2014,bib:mendonca_pop_2014}, it is not yet possible to control the angular momentum degrees of freedom in plasma accelerators. Accessing them is, however, interesting from a fundamental perspective and important for applications, such as radiation generation, which often rely on transverse beam phase space features. %So far, it is only possible to harness the orbital angular momentum of quantum matter wave-packets in non-relsativistic regimes. 

Here, using %a combination of 
theory and particle-in-cell (PIC) simulations, we propose to control the angular momentum degrees of freedom of relativistic beams by introducing the concept of helical-beam plasma wakefield accelerators. The helical-beam wakefield accelerator relies on plasma wakefields with Orbital Angular Momentum (OAM). %Using a combination of theory and particle-in-cell (PIC) simulations, 
We show that %a
%then propose a %fundamentally new type of accelerator: a 
%twisted ,plasma accelerator characterised by non-planar plasma waves with Orbital Angular Momentum (OAM)  where the plasma wave-vector has an additional azimuthal component whose direction varies across the wakefield. We show that the twisted plasma accelerator 
twisted wakefields %
can generate and accelerate relativistic vortex beams, which, unexpectedly, carry quantized levels of angular momentum. %Although their existence was not recognised before, 
%These vortex beams may interact with matter and light in ways not directly accessible to conventional Gaussian beams, for instance, providing access to new degrees of freedom of the radiation that they generate. 
The twisted wakefields could be excited by laser pulses or particle beams with helical profiles. Here, we focus on the laser wakefield accelerator scenario, excited by spatio-temporally shaped femtosecond laser beams called light springs~\cite{bib:pariente_ol_2015} (LS). Light springs have a helical %spatiotemporal 
intensity profile, carry OAM and can effectively transfer %their
angular momentum to the plasma wave. We demonstrate that the LS rotates as it propagates through the plasma, providing an all-optical mechanisms to control the wakefield phase velocity, prolong dephasing and enhance the energy gain in comparison to a planar wakefield. 

%Our work shows that the wakefield topology is an intrinsic degree of freedom of plasma accelerators, which can be exploited to generate relativistic beams with unprecedented properties, beyond current possibilities, thereby significantly expanding the potential of plasma based acceleration.

%We show that, unlike simple LG laser drivers, such beams can transfer their OAM to relativistic plasma waves excited in underdense plasmas, thus radically changing the topology of these waves.

%In the last two decades, LG laser beams have been used in different configurations to transfer OAM to matter~\cite{bib:padgett_nphoton_2011}. This is however not possible for underdense plasmas interacting with standard ultrashort LG laser pulses, 

\begin{figure}
\centering\includegraphics[width= .9\columnwidth]{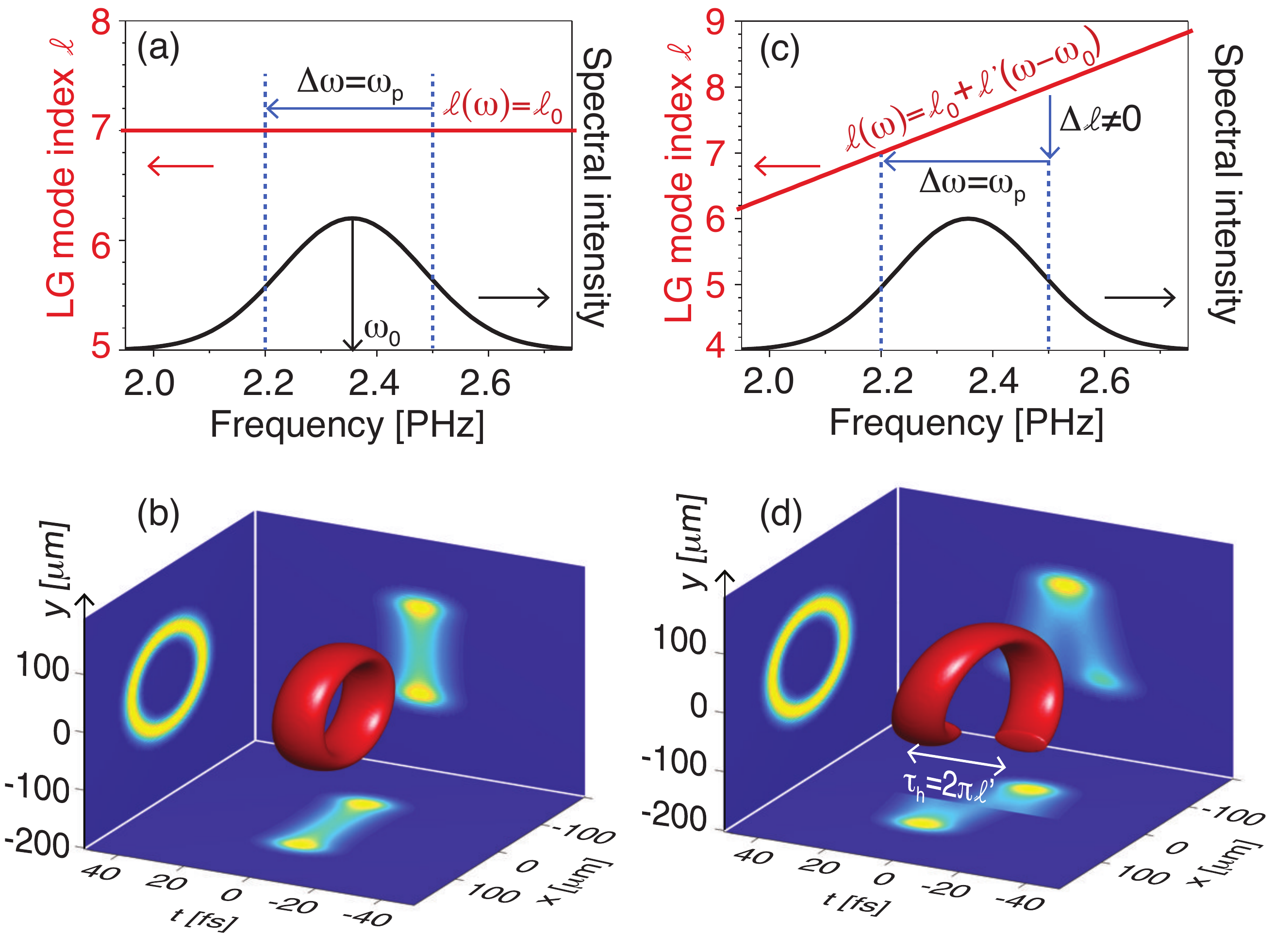}
\caption{Pulsed LG beams (a-b) versus light springs (c-d). The upper graphs show the frequency-dependence of the azimuthal  mode index $\ell$ in the two cases. The lower graphs display the corresponding spatiotemporal intensity profiles of the pulses.}
\label{fig:figure0}
\end{figure}

A pure Laguerre-Gaussian (LG) laser beam with OAM~\cite{bib:allen_pra_1992} cannot transfer its OAM to a plasma wakefield because the %elementary 
mechanism underlying wakefield excitation %the excitation of plasma waves 
is stimulated Raman scattering~\cite{bib:cohen_prl_1972}:  since the absorbed and emitted photons each carry the same OAM $\ell_0 \hbar$, %this does not lead to any 
there is no %
net transfer of OAM to the excited medium [Fig.\ref{fig:figure0}(a)-(b)]. Such a transfer %only 
becomes possible when the OAM per photon in the laser driver is frequency-dependent [Fig.\ref{fig:figure0}(c)]: this corresponds to a spatiospectrally coupled laser beam, where each frequency %in the pulse 
is associated to a spatial LG mode with a different azimuthal index, $\ell=\ell(\omega)$. When $\ell(\omega)$ is linear, such a superposition of modes forms a LS [Fig.\ref{fig:figure0}(d)]~\cite{bib:pariente_ol_2015}. %beam with a helical intensity profile This time-domain perspective makes it clear why these LS can potentially transfer OAM to underdense plasmas: while the OAM of LG beams is exclusively encoded in their wavefronts, the OAM carried by LS is also encoded in their intensity profile, which is the relevant physical quantity for the excitation of plasma waves through the laser ponderomotive force.

%For any laser driver, resonantly exciting plasma waves by stimulated Raman scattering requires pairs of photons separated in frequency by $\Delta \omega=\omega_p$, with $\omega_p=(n e^2/m \epsilon_0)^{1/2}$ the plasma frequency ($n$ plasma density, $e$ and $m$ electron charge and mass, $\epsilon_0$ vacuum dielectric constant). In the case of LS, an additional resonance condition is %intuitively expected 
%required %
%to allow for the transfer of OAM to the plasma: the OAM difference between these two photons (in units of $\hbar$), 
Efficient OAM transfer from the driver to the plasma wake %then
requires that the OAM difference ($\Delta \ell$) between two photons separated by the plasma frequency $\omega_p$, given by $\Delta \ell=\ell(\omega+\omega_p)-\ell(\omega)=d \ell /d \omega \times \omega_p$, %needs to be 
is %
an integer, which then corresponds to the %elementary 
OAM, $\ell_p \in \mathbb{Z}$, acquired by the wakefield. %plasma wave. % for each elementary Raman process. 
This condition %thus
fixes the slope $\ell'=d \ell /d \omega $ of $\ell(\omega)$ to $\ell'=\ell_p/\omega_p$. In the time domain, the physical meaning of this condition is that the temporal pitch $\tau_h$ of the LS intensity helix, given by $\tau_h=2 \pi \left|\ell'\right|$~\cite{bib:pariente_ol_2015}, needs to be an integer multiple of the plasma wave temporal period $\tau_p=2\pi/\omega_p$, to ensure that the plasma wave excited by the final edge of the LS is in phase with the one previously excited by its starting edge.

\begin{figure}
\centering\includegraphics[width=.9 \columnwidth]{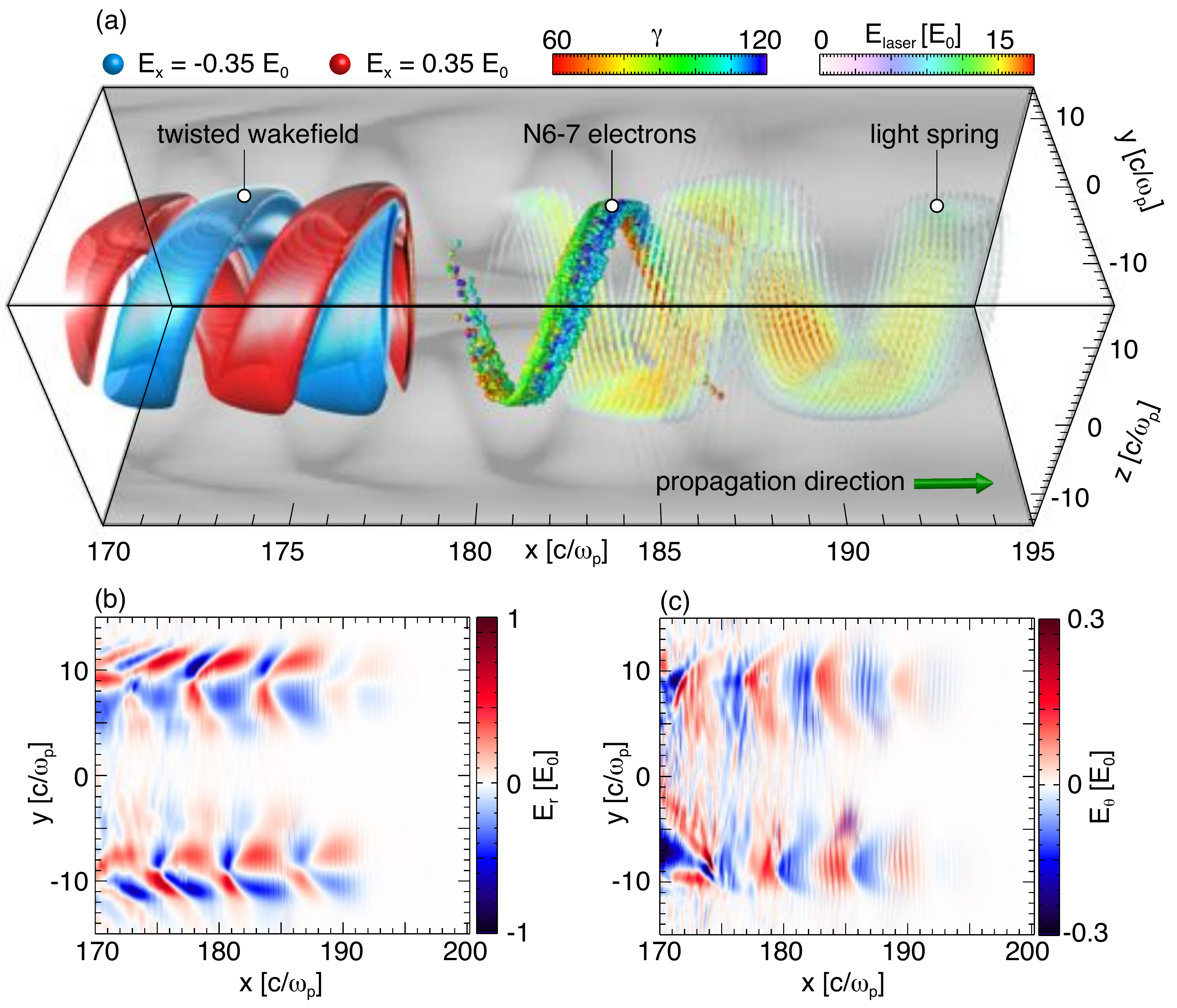}
\caption{Twisted plasma wakefields driven by a light spring moving in a preformed plasma doped with Nitrogen. (a) (right) Rainbow colors display the electric field of the light spring. (left) blue-red isosurfaces show the twisted longitudinal electric field structure of the wake excited by this LS in the underdense plasma. These surfaces are not displayed for $x \geq 178$ to avoid hiding the other plots. (middle) Spheres in rainbow colors correspond to ionization injected electrons from the inner ($6^{th}-7^{th}$) shells of Nitrogen. (b)-(c) slices of the corresponding radial and azimuthal electric fields in the plasma. Field values are normalised to the cold wavebreaking limit $E_0=m_e c \omega_p/e$.}
\label{fig:figure1}
\end{figure}

Figure~\ref{fig:figure1}a shows the results of three-dimensional PIC simulations performed with %the 
Osiris%code
~\cite{bib:fonseca_book,bib:osiris}, illustrating a twisted plasma wave driven by a LS in these resonant conditions, with $\ell_p=1$. The simulation considers a preformed parabolic plasma %density profile 
channel %
to ensure stable propagation. The laser driver %is based on the simplest LS model, which 
incorporates two different LG modes, %only,
with OAM levels differing by $\Delta \ell = 1$ and frequencies (%resp. 
wavenumbers) by $\Delta \omega = \omega_p$ (%resp. 
$\Delta k = k_p$), such that $\ell^{\prime} = 1/\omega_p$. The peak normalized vector potential of each mode is $a_0=0.75$ ($a_0$ %is the normalised laser vector potential, which 
relates to the peak laser intensity through $a_0 = 8.6\times10^{-10}\lambda [\mu\mathrm{m}] I^{1/2} [\mathrm{W/cm^2}]$). In addition a moving window with dimensions $36\times40\times40 (c/\omega_p)^3$ and %divided into 
$2700\times200\times200$ cells was used. Each cell contains 10 particles (see online supplementary material for additional details). Because $\Delta \omega = \omega_p$ ($\Delta k = k_p$) and $\Delta \ell = 1$, this LS [green-yellow-red colors in Fig.~\ref{fig:figure1}(a)] is expected to excite a twisted wakefield with $\ell_p = 1$. This %is indeed what is observed in 
agrees with %
the simulation, for instance by considering the wake longitudinal electric field [blue and red surfaces in Fig.~\ref{fig:figure1}(a)]. More generally, simulations show that when $\ell'$ is gradually increased, field structures with $\ell_p$ helical strands appear in the plasma when the resonance condition $\ell'=\ell_p/\omega_p$ is fulfilled.

The field structure of a twisted wakefield, similar to that shown in Fig.~\ref{fig:figure1}a, can be captured analytically by a wake potential of the form $\phi = \phi_0(r)\cos(k_p \xi + \ell_{p} \theta)$, where $\phi_0(r)$ is the amplitude, $r$ is the distance to the axis, $\xi = x - v_{\phi} t$ is the comoving frame variable with $v_{\phi}\simeq c$ being the wake phase velocity and $\theta$ the azimutal angle. The wake longitudinal electric field is then $E_x = -\partial \phi / \partial \xi = k_p \phi_0(r) \sin(k_p \xi + \ell_p \theta)$. The OAM of the wake potential and electric fields thus coincide. Figure~\ref{fig:figure1}(a) and the analytical expression for $E_x$ also show that a closed path along the $\theta$ direction (for a fixed $\xi$) crosses positive and negative field regions. This contrasts with a non-OAM wakefield where positive and negative fields can only be accessed by following a path along $\xi$. This feature results from the modified topology of the plasma wave: while areas of constant field sign consist of a succession of totally disjoint 'bubbles' in the wakefields created by standard Gaussian beams, here these areas form a set of intertwined helices, with a single continuous helix for each field value (Fig.~\ref{fig:figure1}a).

In contrast with planar wakefields, where the %plasma
wave-vector is constant and given by $k_p\mathbf{e}_x$, the twisted plasma wave-vector is angle dependent, and given by $k_p\mathbf{e}_x + \ell_{p}/r \mathbf{e}_{\theta}$ ($\mathbf{e}_x$ and $\mathbf{e}_{\theta}$ are the unit vectors along $x$  and $\theta$ respectively). This allows exploring new acceleration regimes and generate new types of particle bunches. %
To investigate the new degrees of freedom that now become available, we consider the Panofsky-Wenzel theorem~\cite{bib:panofsky_rsi_1956}, $\nabla_{\perp} E_x = \partial \mathbf{W}_{\perp} /\partial \xi$, with $\mathbf{W}_{\perp} = \mathbf{E}_{\perp}+c\mathbf{e}_x \times \mathbf{B}_{\perp}$, which relates the longitudinal and transverse wakefield components acting on a relativistic particle moving at $c$. Here, $\mathbf{E}_{\perp}$ and $\mathbf{B}_{\perp}$ are the transverse electric and magnetic wakefields, respectively. %As in a planar wakefield, 
In addition to the radial focusing force, which appears since $E_x$ depends on $r$ (Fig.~\ref{fig:figure1}(b)), 
%shows the radial wake electric field component $E_r$, which is responsible for the betatron oscillations of trapped particles.
the Panofsky-Wenzel theorem predicts a new azimuthal field component (Fig.~\ref{fig:figure1}(c)), given by $(1/r)\partial E_{x}/\partial \theta = \partial (E_{\theta} + B_r)/\partial \xi$, given by $E_{\theta}+B_r = \phi_0(r) (\ell_p/r) \sin(k_p \xi+\ell_p \theta)$. The new azimuthal wakefield component is a remarkable feature of twisted plasma waves, which strongly affects the dynamics of background plasma electrons and the dynamics of relativistic trapped particles.
%Figure~\ref{fig:figure1}(c) shows the $E_{\theta}$ wake component, which causes particles to move along $\mathbf{e}_{\theta}$. 

%The new azimuthal wakefield component is %again 
%a remarkable feature of %these helical 
%twisted %
%plasma waves, which strongly affects not only the dynamics of background plasma electrons, but also the dynamics of relativistic trapped particles. %the particle dynamics in the wake. 
%%We now consider the dynamics of electrons accelerated to relativistic energies. 
%To clearly isolate the trapped particles from the background plasma, the simulation of Fig.~\ref{fig:figure1} considers a small concentration of Nitrogen, which allows for the occurrence of ionization injection from the inner $6^{\mathrm{th}}-7^{\mathrm{th}}$ atomic shells of the Nitrogen. A helical particle bunch [rainbow colored spheres in Fig.~\ref{fig:figure1}(a)], with a single helical strand, then forms and %gets accelerated
%accelerates to relativistic energies in the peculiar wakefield structure described above, combining longitudinal, radial and azimuthal components. 

%
The existence of a finite azimuthal wakefield, which is a result of the twisted wakefield topology, has far reaching consequences for the plasma dynamics and acceleration. % 
Unlike planar wakefields, the longitudinal and azimuthal trajectories of the bunch particles are no longer independent. Their relation can be determined using Hamilton's equations. In an % purely
electrostatic wakefield, the canonical momentum along $x$ ($\mathcal{P}_x$) and $\theta$ ($\mathcal{P}_{\theta}$) correspond to the longitudinal (linear) momentum ($p_x$) and angular momentum ($L_x$), given by $\mathcal{P}_x = p_x = m_e c \gamma \beta_x$ and $\mathcal{P}_{\theta} = L_x = r p_{\theta} = m_e c r \gamma \beta_{\theta}$, respectively. In these expressions, $\beta_x$ and  $\beta_{\theta}$ are the longitudinal and azimuthal velocity components normalized to $c$. Hamilton's equations then read $\mathrm{d}_t p_x = -\partial_x \mathcal{H} $ and $\mathrm{d}_t L_x = -\partial_{\theta} \mathcal{H}$, where $\mathcal{H} = m_e c^2 \gamma + e \phi(r,\theta,\xi)$ is the Hamiltonian of a particle with charge $e$ and relativistic factor $\gamma$. Supplementing these equations with $\mathrm{d}_t \mathcal{H} = \partial_t\mathcal{H}$ then leads to: 
\begin{equation}
\label{eq:ebeamoam}
\frac{\Delta L_x}{\Delta p_x} = \frac{\ell_p}{k_p}
\end{equation}
Equation~(\ref{eq:ebeamoam}) is a key result of the work.  %, which now has an additional azimuthal component, and has no parallel in a planar wakefield.
By establishing a proportionality relation between the angular and longitudinal momentum of the accelerated particles, Eq.~(\ref{eq:ebeamoam}) not only affects the dynamics of background plasma electrons, but has profound consequences on the dynamics of relativistic trapped particles, which are evident in the simulations. This proportionality relation is imposed by the wakefield topology. To clearly isolate the trapped particles, %from the background plasma, 
the simulation of Fig.~\ref{fig:figure1} considers a small concentration of Nitrogen, which allows for the occurrence of ionization injection from the inner $6^{\mathrm{th}}-7^{\mathrm{th}}$ atomic shells of the Nitrogen. A helical particle bunch [rainbow colored spheres in Fig.~\ref{fig:figure1}(a)], with a single helical strand, then forms and %gets accelerated
accelerates to relativistic energies.% in the peculiar wakefield structure described above, combining longitudinal, radial and azimuthal components. 

\begin{figure}
\centering\includegraphics[width=.9 \columnwidth]{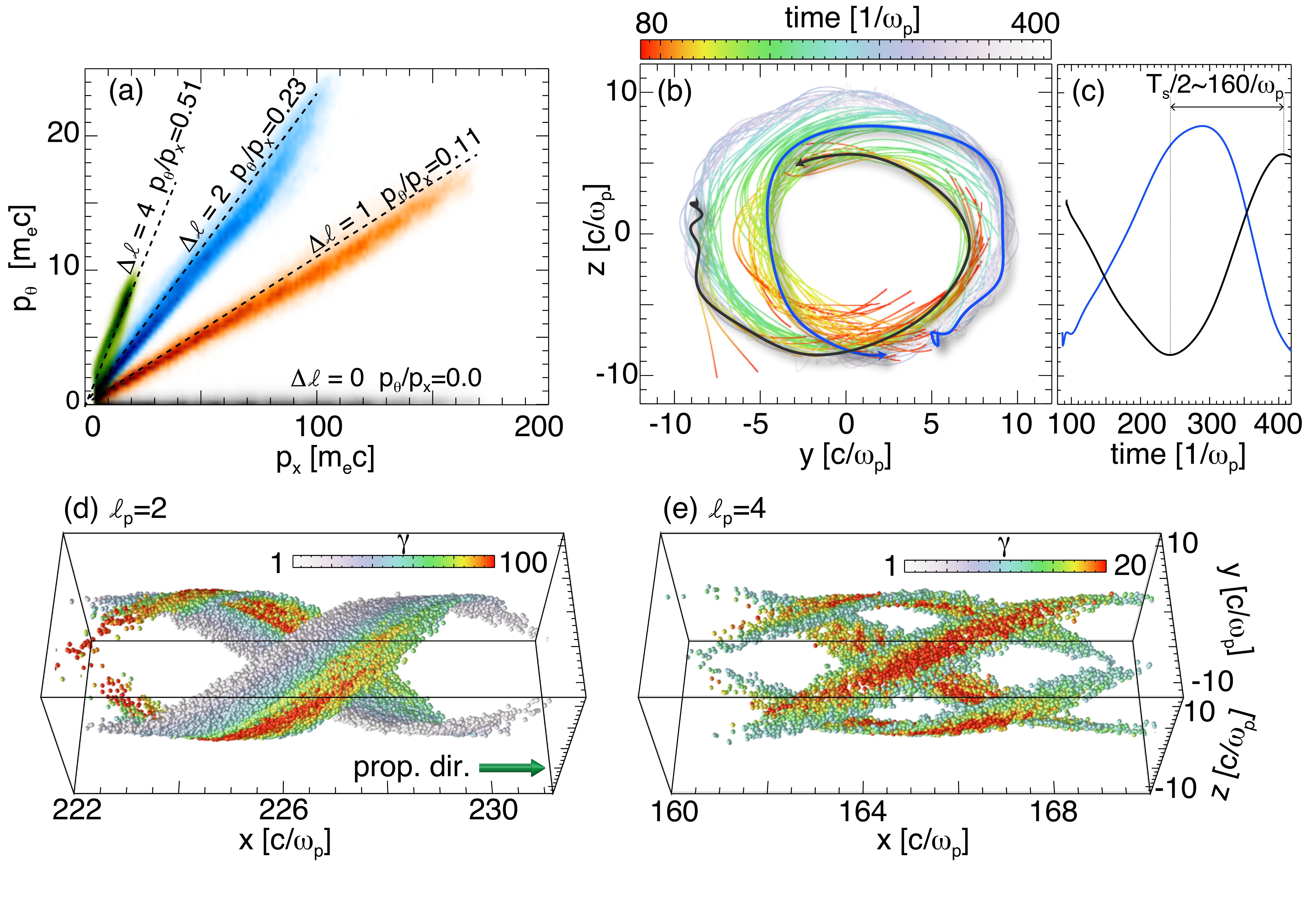}
\caption{Properties of relativistic vortex beams accelerated by twisted plasma waves. (a) $p_{\theta}-p_x$ phase space of vortex electron bunches in plasma waves with different OAM levels. The dashed lines show the prediction of Eq.~(\ref{eq:ebeamoam}) in each case. Note the negligible azimuthal momentum of trapped particles in the case where $\Delta \ell = 0$ (gray distribution). (b) helical trajectories of a random sample of ionization injected Nitrogen electrons forming a vortex beam with $\ell_p = 1$. The trajectories are colored according to the propagation time. (c) trajectories of two electrons [shown in blue and red in (b)] along the propagation time. (d)-(e) three-dimensional vortex particle bunches coloured according to the energy.}
\label{fig:figure2}
\end{figure}

According to Eq.~(\ref{eq:ebeamoam}), for a given $\Delta p_x$, the angular momentum $\Delta L_x$ can only vary in jumps, multiples of $\ell_p$ at a fixed plasma density. Fig.~\ref{fig:figure2}(a) shows the bunch particle distribution in the $p_{\theta}-p_x$ phase-space, obtained from numerical simulation performed with different values of $\Delta \ell=\ell_p$. The slope of these particle distribution varies in discrete steps that indeed follow the theoretical prediction, given by $p_{\theta}/p_x = \ell_p / (k_p r)$. Here, $r\simeq r_0 = w_0 \sqrt{ |\ell_0| /2}$ is the radius where the LS intensity is maximum, with $\ell_0=\langle\ell(\omega)\rangle$ the frequency-averaged value of $\ell$. The azimuthal particle motion then dominates the transverse particle dynamics because $p_{\theta} \gg p_r$. Twisted wakefields can also be exploited to generate beams characterized by comparable longitudinal and transverse momenta ($p_x\sim p_{\theta}$) when $\ell_p \simeq r_0 k_p$. Figure~\ref{fig:figure2}(a) (green region) illustrates this regime when $\ell_p=4$, for which $p_{\theta}\simeq p_x/2$.

A remarkable consequence %of the new twisted wakefield topology
that follows from the quantization of the angular momentum expressed by Eq.~(\ref{eq:ebeamoam}) is that %the azimuthal velocity 
$\beta_{\theta}$ of trapped particles can only take a discrete set of values. The emergence of this unusual quantization rule can be directly linked to the twisted wakefield topology. In the limit where $\beta_x = v_x/c\simeq 1$ and the particle was initially at rest, $\beta_{\theta} \simeq p_{\theta}/p_x = \ell_p/(k_p r_0)$. Thus, $\beta_{\theta}$ is quantized because it can only change by a multiple of $1/(k_p r_0)$, as $\ell_p$ varies. The well-defined slopes of the phase-space regions in Fig.~\ref{fig:figure2}a already reflect this quantization rule. Figure~\ref{fig:figure2}b and \ref{fig:figure2}c further confirm that trapped electrons perform helical trajectories with a period that agrees with theory. Figure~\ref{fig:figure2}b shows the trajectories of a random sample of trapped Nitrogen electrons in the transverse $y-z$ plane for a twisted wakefield with $\ell_p=1$. Figure~\ref{fig:figure2}(c) shows that the spiralling period is $T_s \omega_p \simeq 320$, close to the theoretical estimate, given by $T_s \omega_p = 2 \pi k_p^2 r_0^2 / \ell_p$, which yields $T_s \omega_p \simeq 353$.

Although purely classical, this quantization rule for $\beta_{\theta}$ is analogous to the OAM quantization of twisted rays of light as well as quantum vortex electron wavepackets. To establish a parallel with twisted light, we consider the paraxial approximation, where $\mathbf{p}_{\perp}\ll p_x$, so that the energy ($E_p$) of a relativistic particle is $E_p \simeq c p_x \simeq \gamma m_e c^2$. Hence, Eq.~(\ref{eq:ebeamoam}) leads to $L_{x}/E_p \simeq \ell_p/\omega_p$. The latter expression is analogous to the quantization of the ratio of angular momentum flux to energy flux for a LG beam~\cite{bib:allen_pra_1992}, given by $M_z = \ell/\omega$. It also recovers the OAM quantization of vortex free electron quantum wave packet, given by $L_x/p_x = \ell/k_x$, where $k_x$ is the wavepacket wave-number~\cite{bib:bliokh_prx_2012}. These similarities suggest that trapped particles may be seen as a matter analogue of an OAM light beam and a classical matter analogue of a quantum electron wavepacket.

%To show that the accelerated particles form a true vortex beam, their spatial distribution needs to be helical, with a well defined number of helixes given by $\ell_p$, consistent with the angular momentum quantization rule. Figures~\ref{fig:figure2}(d)-(e), together with Fig.2(a), demonstrate this hypothesis. 

Figure~\ref{fig:figure2}(d), which illustrates the density distribution of the beam with $\ell_p = 2$ [blue region in Fig.~\ref{fig:figure2}(a)] and Fig.~\ref{fig:figure2}(e) with $\ell_p=4$ [green region in Fig.~\ref{fig:figure2}(a)], shows the correspondence between the spatial and velocity distribution of the relativistic bunches: the number of helixes, given by $\ell_p$, is proportional to the angular momentum, given by the quantization rule $p_{\theta}/p_x = \ell_p/(k_p r)$, thus demonstrating that these beams constitute a new type of charged particle beams %whose existence has remained unnoticed: classical vortex bunches with OAM.
with a vortex spatial structure and with quantised levels of angular momentum.

A general feature of the acceleration in a twisted wake is that particles can also dephase along $\mathbf{e}_{\theta}$, thereby potentially lowering the dephasing length and energy gain in comparison to a planar wakefield. Figure~\ref{fig:figure2}(a) confirms this prediction, showing that the energy gain decreases for larger $\Delta \ell$. This effect can be mitigated due to a new and intrinsic property of twisted wakefields: as a LS propagates in the plasma, its intensity helix rotates azimuthally. This rotation enables to adjust the wakefield phase velocity all-optically. Remarkably, in some cases, the acceleration can even become stronger than in a planar wakefield.

\begin{figure}
\centering\includegraphics[width=.9 \columnwidth]{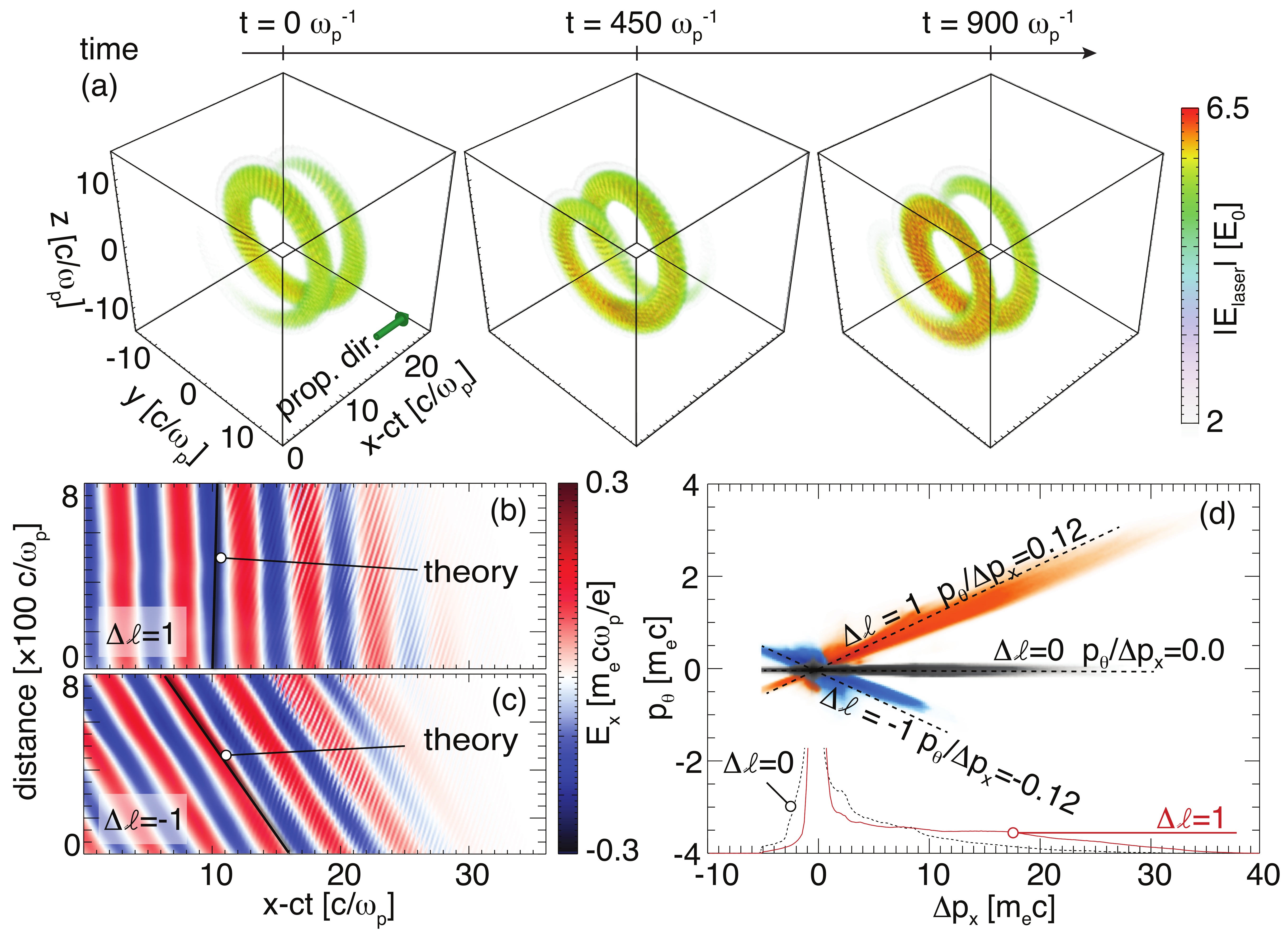}
\caption{Effects of the LS azimuthal rotation. (a) Spatiotemporal intensity profile of a LS propagating in a plasma channel, at different times in the propagation.  (b)-(c) Evolution of the accelerating fields at $y\simeq 9~c/\omega_p$ and $z=0$, corresponding to the region around the radial maximum of the LS intensity, for a fixed $\ell_0>0$. The theoretical predictions of Eq.(\ref{eq:groupvelocity}) are shown as black lines. In panel (b), the phase velocity of the wakefield is very close to $c>v_g$, due to the effect of the LS rotation. (d) $p_{\theta}-\Delta p_x$ phase space in twisted plasma waves with different OAM levels. The theoretical predictions of Eq.~(\ref{eq:ebeamoam}) are given by the black dashed lines. Solid red and point black curves respectively show the spectra of accelerated particles.}
\label{fig:figure3}
\end{figure}

While signatures for this rotation can be observed in a uniform plasma in the presence of self-guiding induced by the non-linear plasma response, this effect becomes particularly clear, and its derivation simpler, in the presence of a preformed parabolic plasma channel, in the linear propagation regime. Figure~\ref{fig:figure3}(a) illustrates this azimuthal rotation of the LS intensity helix, as it propagates in such a preformed plasma channel (see online supplementary material for the simulation details).

The channel counteracts diffraction by preventing the wavefronts to curve outwards. The angular momentum of a twisted light ray is preserved in the channel, causing the LS intensity profile to spin around its axis, in a direction determined by the helicity of the wavefronts, i.e. by the sign of $\ell_0$. Just as in a rotating screw, the laser pulse at a given radial position appears to move either backward or forward in the frame of the laser, depending whether this ray rotation is opposite to, or along the LS intensity helix, which direction is determined by the sign of $\ell'=d\ell/d\omega$. Thus, when looking at a given longitudinal plane (along $x$ for fixed $y$ or $z$), the laser pulse envelope will appear to move either slower or faster than the linear group velocity in the plasma, depending on the relative sign of $\ell_0$ and $\ell'$.

The resulting effective group velocity can be determined analytically (see online supplementary material for an explicit derivation) is: 
\begin{equation}
\label{eq:groupvelocity}
v_{g, \mathrm{LS}} =  v_{g} + \frac{2 c^2}{k w_0^2} \ell' \sign\left(\ell_0\right),
\end{equation} 
where $ v_{g}$ is the usual group velocity in the plasma.

The plasma responds to this effective local group velocity of the LS. Hence, the correction term resulting from the rotation of the LS can be exploited to increase or decrease the wakefield phase velocity. Figure~\ref{fig:figure3}(b)-(c) show the evolution of the wakefields as a function of propagation distance when $\ell^{\prime} >0$ [Fig.~\ref{fig:figure3}(b)] and $\ell^{\prime} <0$ [Fig.~\ref{fig:figure3}(c)], for a fixed $\ell_0>0$, and confirm this prediction. This effect vanishes when $\ell^{\prime} = 0$, and is thus an unique feature of twisted wakefields driven by the LS.
 
Since the energy gain of a group of externally injected particles depends on the wake phase velocity, this effect should modify the spectra of particles accelerated by twisted wakefields. This is illustrated on Figure~\ref{fig:figure3}(d): the accelerated bunch reaches higher energies when $\ell'$ and $\ell_0$ have the same sign (orange region), than when they are of opposite signs (blue region).  For these parameters (i.e. keeping laser energy constant in all cases), Fig.~\ref{fig:figure3}(d) also demonstrates that the maximum energy and number of accelerated particles in the twisted wakefield (red solid line) exceeds those produced in a planar wakefield excited by a standard LG beam of same $\ell_0$ (black dashed line).

Most physical schemes studied so far to transfer angular momentum from lasers to plasmas have relied on circularly-polarized light, i.e. on the spin angular momentum of light~\cite{bib:kostyukov_pop_2002,bib:shvets_pre_2002,bib:naseri_pop_2010,bib:liseykina_njp_2016,bib:lecz_lpb_2016,bib:yu_prl_2013,bib:tamburini_pre_2012}, leading to strong magnetic fields, intense radiation bursts, %
and helical beam-plasma structures in dense plasmas exposed to superintense laser fields in the context of radiation pressure dominant acceleration~\cite{bib:tamburini_pre_2012}.
In contrast, our work describes a new configuration where it is now the orbital angular momentum of a laser field that is transferred to an underdense plasma. This enables control of the wakefield topology, which can then be exploited to generate relativistic beams with unprecedented properties, beyond current possibilities, thereby significantly expanding the potential of plasma based acceleration. This configuration can have broad implications in different fields~\cite{bib:vieira_prl_2014,bib:denoeud_prl_2017,bib:leblanc_natphys_2016,bib:brabetz_pop_2015,bib:zhang_prl_2015,bib:baumann_qe_2017,bib:izquierdo_np_2016,bib:vieira_natcomms_2016,bib:vieira_prl_2016}, from the generation of twisted x-rays and intense magnetic field generation in plasmas~\cite{bib:shi_submitted,bib:nuter_arxiv_2018} to novel pathways to control laser-matter interactions, and perhaps manipulate the spin in compact plasma based spin-polarisers (e.g. due to the Sokolov-Ternov effect~\cite{bib:sokolov_1963} or spin-precession~\cite{bib:vieira_prstab_2011}). The required spatio-temporally shaped ultrashort laser beams can in principle already be obtained experimentally with simple helical mirrors such as those presented in Ref. \cite{bib:Ghai_AO_2011}. Looking further ahead, advances in ultrafast optical metrology \cite{bib:Pariente_2016} and shaping \cite{bib:Sun_2015} should soon provide access to advanced and programmable spatio-temporal control of ultrashort laser beams, and thus make it possible to tailor the topology of laser-plasma accelerators and hence access new intrinsic degrees of freedom of the resulting high-energy particle beams.

\begin{acknowledgments}
We acknowledge the grant of computing time by the Leibnitz Research Center on SuperMUC. Work partially supported by the EU ARIES Grant Agreement No 730871 (H2020-INFRAIA-2016-1). J. V. acknowledges the support of FCT (Portugal) Grant No. SFRH/IF/01635/2015, and F.Q. the support from the European Research Council under the European Union's Horizon 2020 research and innovation programme (ERC grant agreement 694596).
\end{acknowledgments}

\end{document}